\documentclass[conference]{IEEEtran}
\IEEEoverridecommandlockouts
\usepackage{cite}
\usepackage{amsmath,amssymb,amsfonts}
\usepackage{algorithmic}
\usepackage{graphicx}
\usepackage{textcomp}
\graphicspath{{images/}}
\def\BibTeX{{\rm B\kern-.05em{\sc i\kern-.025em b}\kern-.08em
    T\kern-.1667em\lower.7ex\hbox{E}\kern-.125emX}}

\begin{document}

\title{Internet Protocol Version 6: Dead or Alive?\\
\thanks{}
}

\author{
  Sumit Maheshwari\\
  \textit{WINLAB/ECE, Rutgers University} \\
  \texttt{sumitm@winlab.rutgers.edu}
  \and
  Richard P. Martin\\
  \textit{WINLAB/CS, Rutgers University} \\
  \texttt{rmartin@scarletmail.rutgers.edu}
}

\maketitle

\begin{abstract}
Internet Protocol (IP) is the narrow waist of multi-layered Internet protocol stack which defines the rules for data sent across networks. IPv4 is the fourth version of IP and first commercially available for deployment set by ARPANET in 1983 which is a 32 bit long address and can support up to ${{2}^{32}}$ devices. In April 2017, all Regional Internet Registries (RIRs) confirmed that IPv4 addresses are exhausted and cannot be allocated anymore implying any new organization requesting a block of Internet addresses will be allocated IPv6. This creates troubles of interoperability, migration and deployment, and therefore organizations hesitated to use IPv6 borrowing IPv4 addresses from other big organizations instead. Currently, when IPv4 is not available, and IPv6 is not adopted for around 20 years, the question arises whether IPv6 will still be accepted by the computer society or will it have an end of life soon with alternate better protocol such as ID based networks taking its place. This paper claims that IPv6 has lost its deployment window and can be safely skipped when new ID based protocols are available which not only have simple interoperability, deployment and migration guidelines but also provide advanced features as compared to IPv6. The paper provides answers to these questions with a comprehensive comparison of IPv6 with its available alternatives and reasons of IPv6 failures in its adoption. Finally, the paper declares IPv6 as a dead protocol and suggests to use newer available protocols in future.
\end{abstract}


\section{Introduction} \label{intro}
Internet is a complex network of networks which require ways to identify connections. A connection may be a session between two clients (peer to peer), a client and server, between two routers and so on. Each of these connections are required to be uniquely identified by the network for example a client and server session is unique in the four tuples of \textless source IP, destination IP, source port, destination port\textgreater. In this context, Internet Protocol (IP) is an essential layer in communication stack which precisely provides a location or network address of a connected device. 
In general, each device needs to have an IP address at a particular location which changes when the UE (user equipment) moves and connects to another access point or base station.\par
Internet as we know of today is a completely modified version of what was envisioned during its inception in 1970s where access was designed for fixed users rather than mobile users. This design reflects in the IPv4 as well where IP address changes for a mobile user by means of a sophisticated mechanism of transferring connections from one access point to another which we call handover. As the size of Internet is growing, IPv4 which is a 32 bit IP address is unable to accommodate all the users and therefore a transition towards a 128-bit long IPv6 address is being sought. While IPv6 is a much improved version of IPv4 with additional features such as security, large address space, new header format, stateless and stateful address configuration and so on, it still relies upon the basic mechanisms of IP which is not static with respect to a user or device but changes when a user moves. After around 15 years of operation of IPv4, in 1998, Internet Engineering Task Force (IETF) devised a 128-bit long IPv6 address due to IPv4's depleting addressing space for future. IP is used in all the communication and networking devices worldwide including but not limited to the newest Internet-of-Things (IoTs) sensors.\par
During the IPv6 design phase, several key features were not addressed and therefore it did not see the bright light for deployment even after 20 years. One of the major issue for IPv6 deployment is inability to provide backward compatibility. This means for a network to adopt to this sarcastically newer version of IP, it still have to support IPv4 through some proxy or encapsulation means. This incurs additional cost for the network provider as all the switches and routers work with IPv4, and translation require additional boxes to be placed in network which not only are expensive but also introduce delays (which is network cost). Moreover, since IPv6 was not compatible with IPv4 \textit{as-is}, network providers deferred from this expensive addressing space. \par
During IPv6 finalization phase, Network Address Translation (NAT) boxes were already running in enterprise and organizational networks which isolated the need of providing global IP addresses for the users in these places and therefore, partially, the IPv4 address space was extended indirectly which demotivated network providers although everyone knew that NAT cannot scale well and ultimately they will have to think of alternatives. \par
\begin{figure}[t!]
\begin{center}
\vspace{-1mm}
\includegraphics[width=\linewidth]{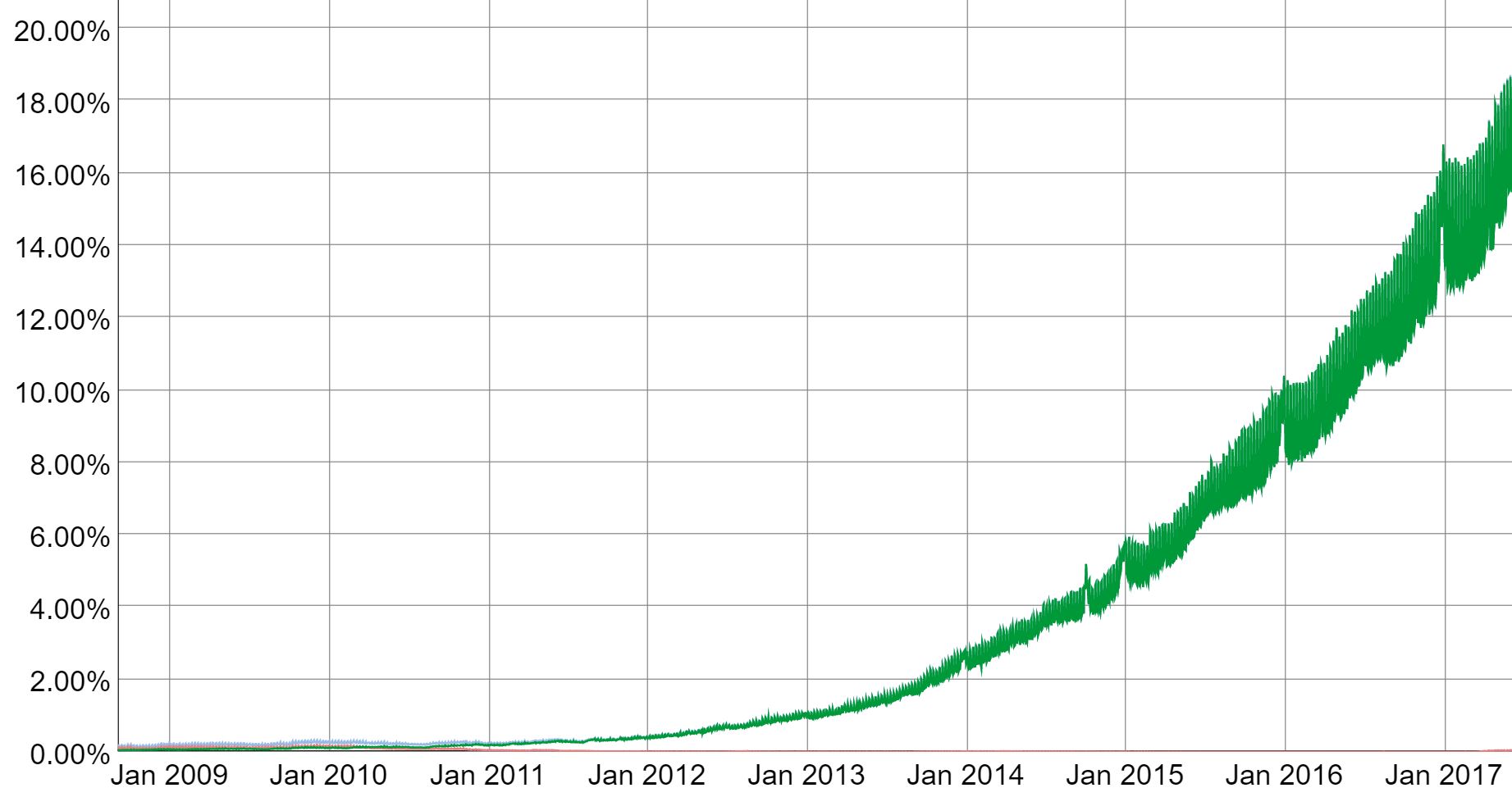}
\vspace{-2mm}
\caption{20 Years IPv6 Deployment Statistics}
\vspace{-4mm}
\label{fig:deployment}
\end{center}
\end{figure}

In general, there was no motivation or incentive for network providers to switch to IPv6 and to save their additional cost, everyone waited till their network partner or peer moved to IPv6 which either never happened or only deferred them. In general, the benefit could be seen only if all the peers would have moved together which is an impractical assumption and classical case of paradigm shift. Although forums such as Internet Society later introduced world IPv6 day \cite{world} which is celebrated every year around the globe on 6th June and was initiated in the year 2002, to bring companies, organizations, home networking device manufacturers and web companies together to support IPv6 by sharing their strategies and progress, even after 15 years of continuous deliberate efforts, IPv6 deployment is not pleasing as only about 20\% of world's total traffic is carried in IPv6 as shown in fig. \ref{fig:deployment} \cite{goog} and 80\% of users are still using IPv4.\par

The questions we ask in this paper are as follows: (a) Is IPv6 the single best networking solution which we are pursuing for last 20 years?, (b) Can IPv6 cater to evolving users' need, which were not even there when it was designed?, (c) Should we provide more efforts to bring remaining 80\% of users to IPv6 or do we have a better alternative?, and (d) Can find a single unified solution which not only serve fixed users for whom the IP protocol was designed or serve the whole community of all kind of users which are highly mobile in today's fast-moving world? This position paper provides answers to these questions while suggesting alternative solutions, particularly claiming that IPv6 is still dead and cannot grow anymore. We provide reasons for why we should assume that IPv6 is dead already and logically seek solutions to this worldwide problem of scalability. \par
The paper is organized as follows. Section II explains  features of IPv6 and modifications from its ancestor. Problems in adopting IPv6 are detailed in Section III. Section IV highlights currently unavailable features of IPv6 and Section V details alternatives of IPv6 for a global transition. Section VI compares the suggested alternative solution to IPv6 while conclusion is made in Section VII.

\section{Features of IPv6}
This section reviews the features of IPv6 in order to gain an overall understanding and build a narrative towards alternative solutions. Following is the list of important features as listed in Request for Comments (RFC) 2460\cite{rfc}:
\begin{itemize}
\item Increased Address Space: As version 4 is unable to support scale of devices, version 6, with its 128 bits address space provides ${{2}^{128}}$ addresses which is a sufficiently large number.
\item Change in Header: Some of the IPv4 headers are dropped to make IPv6 header lightweight, hence saving bandwidth.
\item Enabling Future Options: This new version's header has optional fields for encoding which can be used for efficient forwarding without limiting the length of options. Also, the design is flexible for future use-cases which are unseen.
\item Labeling Flows: Similar to Virtual Networks (VNs) or Multiprotocol Label Switching (MPLS), labeling of flows is featured for fast forwarding.
\item Privacy and Security: Finally, in-built data integrity, and privacy features are designed for a secure forwarding.
\end{itemize}
The reasons of non-deployment of IPv6 are detailed in next section.

\section{IPv6 Adoption Problems}
In this section, we highlight specific problems associated with IPv6 which deferred its deployment for so long in time as highlighted in fig. 1 in section \ref{intro}. In general, with the nice features proposed by Internet Engineering Task Force (IETF) and deliberate efforts by multiple organizations, it was expected that IPv6 will take off. Nonetheless, to learn from this experience, we have listed here issues relating not only with the deployment but also user expectations including but not limited to research data. 
\subsection{Compatibility}
As mentioned in Section \ref{intro}, backward compatibility was one of the major hurdle in implementing IPv6 globally. The restrictions such as both servers and clients have to use same version of IP was a bottleneck and demotivated providers first. The website developers relied upon network providers and vice-versa whereas users were not concerned about this change at all. This led to the concept of \textit{IPv6 only} \cite{only} which is equivalent of creating two parallel networks losing the concept of unified Internet as such.
\subsection{IPv6 as an Overlay}
In order to solve the problem of parallel network, translation services using middleboxes began which incurred additional costs to the network providers and in turn to the users of IPv6. An overlay carries IPv6 traffic either using encapsulation by packaging IPv4 packets inside an IPv6 packet or by translation which discards the headers of IPv4, packaging payload into a new IPv6 packet at the entry point. At the exit of IPv6 network (traffic merging to IPv4), a reverse process is followed. Both of these approaches introduce computational delay or additional bandwidth utilization or both which ultimately translates to the cost.
\subsection{Lack of Connectivity}
Initially, most of the Autonomous Systems (ASes) did not support IPv6 which discouraged small IPv6 network islands' expansion. This lack of connectivity also became a huge problem in deployment \cite{islands}.
\subsection{Fewer Research Data}
There was a serious lack of global research data displaying performance improvement for IPv6 or at least showing equivalence with IPv4 which again demotivated network providers to take initiatives to deploy IPv6. 
\begin{center}
\textit{What is not measured, is not quantified and therefore, is not optimized}
\end{center}
\subsection{Trouble Merging}
For company who wanted to merge to another company by bringing their networks together, also had to think about merging differently running networks (in this case may be IPv4 and IPv6) which was an additional cost to consider to unify the network and therefore small companies deferred to switch to IPv6 till a bigger company did so which created an open loop to deployment.
\subsection{No Peer Pressure}
In the scenario like this, when no one was incentivized to deploy IPv6, all the small network providers were waiting for others to deploy first in the fear of losing their old customers or having to incur additional cost. On the other hand, big network providers had captured huge amount of IPv4 addresses and using techniques such as NAT, were able to support their customers. Therefore, no peer pressure was a bad indicator and further deferred deployment.
\begin{center}
\textit{I would do, if you do, who will begin, first?!}
\end{center}

\section{Lack of IPv6 Features in Emerging Networks}
Since IPv6 was first proposed 20 years back, network has evolved a lot and there are a huge set of features which are to be supported lacking in IPv6. In emerging networks which are becoming more connected with sensors such as Internet of Things (IoTs), communication among vehicles (V2V), interaction of vehicles to infrastructure (V2I) and smart services such as Augmented and Virtual Realities (AR/VR), there is a clear transition from the requirements supported by two decades old network protocol. In case of newer protocols, Following list provide features which are currently not supported by IPv6 inherently and are imposed using overlay or indirect techniques.
\subsection{User Mobility}
Internet was designed for fixed users and does not support mobility inherently. As number of worldwide mobile users have reached 4.7 Billion in 2017 \cite{data1}, and IPv6 does not support them by design, there is a clear need of a protocol which take advantage of user mobility. Techniques such as horizontal and vertical handover provide user mobility at the cost of control switching which in turn changes the IP address of device. With increasing mobility of users with various forms of transportations, this design leads to ever increasing control switching. Also, when a user moves from one Access Point (AP) to another, the network capacity has to ensure that it has enough IP addresses to support incoming users which leads to additional cost and management (for example in Long Term Evolution i.e. LTE, mobility management entity, MME, handles these users).
\subsection{Subscriptions and Multicast}
Current decade has seen multitude of multicast and broadcast application platforms such as Facebook, WhatsApp, WeChat, Line, Instagram, Twitter and so on. The traffic carried in current network relies upon multiple end-to-end sessions from server to client due to the sheer limitation that each device has to have an IP address. This design works well for small number of users in a network while for increase in number of users, there is a lot of redundant data flow in network which only wastes network resources and hence increases the service providers' cost. Although, in some routers \cite{cisco} multicast can be enabled by choice, there are limitations of interoperability and singular operations. This sparks an interesting design choice to support multicast as a network unicast to save bandwidth currently unavailable in IPv6.
\subsection{Multihoming}
Multihoming supports multiple network technologies on a single device (example LTE and Wi-Fi on same mobile). In today's network with IPv4 or IPv6, multihoming is supported only using end to end support which cannot be dynamically used based upon device capabilities. Further, network protocols have little or no role in providing this support and therefore server and client must initiate multiple TCP sessions beforehand to get this benefit. With more and more devices having multiple network interface, the same should be supported by the networks on-the-go which is not provided by IPv6 as such.
\subsection{Failure Handling and Data Reliability}
Failures are common in network which are to be addressed dynamically. In current network, in case of a network failure, which strictly follows a fixed IP format, failure leads to retransmission with network providing no support in storing data intermittently or providing a different IP address for a new connection on the fly. This is again a limitation inbuilt in design of IPv6 and a solution is to be sought for the same.\par
Problems lose their values without corresponding solutions and therefore next section describes available solutions which can provide these features and more.
\subsection{Security}
Traditional levels of security follow a tunneling mechanism such as IPv6Sec which fails integrity tests at multiple network points due to don't fragment bits set. There are also problems when fragmenting such packets which are dropped by intermediate routers assuming these to be invalid packets. These security issues can be easily resolved by newer techniques such as cryptography using ID based networks.

\section{Available Alternatives}
IP is facing criticism in research community due to its inability to deal with what we used to call advanced network features but are very common now due to multitude of applications and use-case emerging out of them such as (a) handling mobility, (b) supporting multihoming, (c) providing inbuilt multicast, and (d) reliably carry data in network. Efforts are made to provide these features in an easy, cost effective way thereby avoiding the need of bringing remaining 80\% of Internet users to IPv6 which does not have these features in design. As the network is transitioning and struggling for address space, several solutions are proposed in literature wherein identity based networks are most commonly recognized and are being supported by industries and academia alike. Below are some of these solutions which rely upon the concept of fixed name instead of dynamic IP addresses and hence supporting these advanced features we discussed.
\subsection{MobilityFirst (MF)}
MobilityFirst (MF) \cite{mf} is one of the five Future Internet Architecture's (FIA) funded by National Science Foundation (NSF) supporting the concept of fixed flat names for devices, group of devices and network entities instead of using IP addresses. A globally unique identity (GUID) is a fixed name associated with a device and does not change with mobility of user. This provides a mechanism to separate names and network addresses which is a single function in IP and needs to continuously change when the user moves. The names and network addresses are mapped in a global name resolution service (GNRS), which is similar to Domain Name Server (DNS), thereby dynamically resolving the device location in case of mobility, multicasting, multihoming and failures without changing its identity. This simple design can be easily implemented using a fixed IP address as name while provide a network address as an additional parameter for each user, stored in a logically centralized, physically distributed database. The GUID with its feature as proposed in the architecture, can be assigned to group of users, routers, switches and even contents thus providing additional features such as privacy, security, self-authentication and anycasting supported inherently. The cost of Dynamic Host Configuration Protocol (DHCP) box can then also be reduced as there will not be a need to assign names/IP address each time a user connects from a different location. While one may argue that this design will face similar deployment issues as IPv6, decrease in network cost due to these additional features would save overall cost for network providers once they switch to ID based network.
\subsection{Named Data Networking (NDN)}
Another NSF funded project, Named Data Networking (NDN) \cite{ndn} originated based on the fact that the Internet is moving from host-centric to data-centric network with emergent Information Centric Networks (ICN). Similar to MF, NDN supports unique global names while in a hierarchical manner (example house/table/lamp/bulb) as compared to flat name as supported by MF (example: mytablelampisthis). Features such as routing, forwarding, trust, security, in-network storage etc. all revolve around names instead of using IP hence supporting advance services which are not supported by IPv6 as such. Again, the cost-effectiveness of the solution should be the main benefit for network providers which is unavailable in IPv6 deployment. 

\subsection{Host Identity Protocol (HIP)}
Another interesting concept, Host Identity Protocol (HIP) \cite{host} cleverly defines identity as the unique property of the device (example name) while identifier is the network address of the device, thereby distinguishing name from the current location of name itself. HIP is again a name based RFC of Network Working Group (NWG) and is being considered as a draft to define future of networking without IP addresses. HIP defines a new namespace, Host Identity (HI) and Host Identity Tag (HIT) which are similar to public key and its hash respectively for additional security purposes. With 128 bits HIT, similar to IPv6 addressing, HIP is simple to implement and deploy while providing additional benefits to the network provider as they can handle mobility, security, multihoming, multicasting and so on inherently in design. Also, the self-certifying nature of HIT provides data integrity and privacy features. \par
On one hand, industry consortium with companies such as Microsoft, Nokia, Huawei etc. supporting the concept of name based networking, while on the other hand, researchers are actively working to make this a feasible solution, we can justify that when compared evenly on the grounds of features vs. cost, IPv6 will be considered dead in future. As ID based networking is becoming popular with researchers deploying it in lab environment for various purposes, it is just matter of time for it to become popular among network providers as well.

\section{Names Overcoming Shortcomings}
This section highlights that how names are next generation of networking irrespective of some feature of IP which cannot be easily replicated in name based networks. IP has some unique advantages which are inherent in its design for instance traffic aggregation and ability to cache content which are missing in name-based networks. Following points define the pursuit of names in light of these features and justify our stand on those.
\subsection{No Aggregation in Names}
While NDN supports hierarchical naming schema which has capability of traffic aggregation similar to IP addressing scheme which localizes traffic using prefixes. For example, in /room/table/lamp/bulbs/bulb1, the traffic can be aggregated at room  level, and then be sent to all the lamps in that room consequently to the bulb1 as well. While this design works in case of single hierarchy, it fails when the same bulb is also a part of an another level of hierarchy for instance /company/Philips/devices/bulbs/bulb1 in which case, it becomes a complex graph as compared to tree structure and is difficult to manage. \par
MF supports flat naming scheme which has no aggregation as such while assigning names. The traffic aggregation is possible if group of users are accessing similar content from a same geographical location. The concept of aggregation arrives from our understanding of fixed users which is diminishing with more users move with high speed nowadays. The same can be understood using a simple mobile phone example as follows. In past, in telecommunications, first 3 digits of a cellphone number were used to identify the location of a user which in today's world no longer valid as the user can be anywhere in the much bigger area such as country or world as whole. Therefore, basestation has to have the capability to page the entire region irrespective of the user's home location stored in Home Location Register (HLR). Taking advantage from this discussion, we believe that with more mobile users, the ability to dynamically locating users make more sense than focusing on the concept of aggregation based upon prefixes which is not a valid scenario anymore. \par
Therefore, irrespective of using flat or hierarchical naming schema, we can see that name based networks have advantages over IP based networks. 
\subsection{Content Caching}
On similar line of thought, content caching using the IP prefixes is an old concept which is outdated by ICNs where the focus is on content and not the user of the content. Once we know that the content is becoming popular by looking at content GUID, we can easily think of an expanded concept of caching as well pre-fetching. The dynamics of such networks, which are becoming content oriented, allows us to use newer techniques of forecasting, ranking, content sharing and so on instead of just relying upon classic IP based content caching. Therefore, the name based networks can serve the need of the hour without affecting user's experience. \par
Finally, with the help of advanced services provided by lively ID based networks, one can achieve much more than waiting for the world to adopt already dead IPv6.

\section{Conclusion}
IP protocol version 6 (IPv6) had been waiting to see the worldwide deployment since two decades. With the emergent networking techniques, application requirements and ample modifications in the network itself, IPv6 cannot support the advanced features such as mobility, multihoming, failure handling and multicasting intuitively and inherently, which were not a part of original design and are being augmented using overlays. Industry and academics have joined hands in researching identity based networks which not only can handle basic IP services but also can provide advanced features unsupported by IPv6. The evolving nature of networks and continuous efforts by the community have proven that separating names from network location of a device can solve complex problems trivially. In this position paper, we discussed reasons of IPv6 deployment failures and provided alternative solutions in place of IPv6. Also, it is justified that it will be cost-effective to use newer solution than to pursue remaining 80\% of users to adopt IPv6. Finally, ID based networking is proved to be a single solution which can cater wide needs of users.

\end{document}